\documentclass[twocolumn,aps,psfig,epsfig,bm,showpacs,superscriptaddress]{revtex4} 
\usepackage{ulem}
\usepackage{epsfig}
\usepackage{color}
\begin{document} 

\newcommand{\kmax}{K} 
\newcommand{\Qp}{Q^{+}_\mu} 
\newcommand{\Qm}{Q^{-}_\mu} 
\newcommand{\Qpk}{Q^{+}_k} 
\newcommand{\Qmk}{Q^{-}_k} 

\newcommand{\Red}[1]{\textcolor[named]{Red}{#1}}
\newcommand{\Blue}[1]{\textcolor[named]{Blue}{#1}}
\newcommand{\Green}[1]{\textcolor[named]{Green}{#1}}

\newcommand{\strike}[1]{\textcolor[named]{Red}{[\sout{#1}]}}
\newcommand{\Redmg}[1]{{\em \bf mg:}\textcolor[named]{Red}{#1}}
\newcommand{\Redmd}[1]{{\em \bf md:}\textcolor[named]{Red}{#1}}
\newcommand{\Redrv}[1]{{\em \bf rv:}\textcolor[named]{Red}{#1}}
\newcommand{\Redsw}[1]{{\em \bf sw:}\textcolor[named]{Red}{#1}}
\newcommand{\mg}[1]{{\em \bf mg:}\textcolor[named]{Blue}{#1}}
\newcommand{\sw}[1]{{\em \bf sw:}\textcolor[named]{Blue}{#1}}
\newcommand{\md}[1]{{\em \bf md:}\textcolor[named]{Blue}{#1}}
\newcommand{\Bluerv}[1]{{\em \bf rv:}\textcolor[named]{Blue}{#1}}
\newcommand{\Bluesw}[1]{{\em \bf sw:}\textcolor[named]{Blue}{#1}}

\newcommand{\Greenmg}[1]{\textcolor[named]{Green}{[{\em \bf mg:}#1]}}
\newcommand{\Greenrv}[1]{\textcolor[named]{Green}{[{\em \bf rv:}#1]}} 

\newcommand{\wh}[1]{{\em \bf wh:}\textcolor[named]{Magenta}{#1}}
\newcommand{\strikew}[1]{\textcolor[named]{Green}{[\sout{#1}]}}
\newcommand{\whspace[1]}{\wh{\underline{ }}}

\newcommand{\vk}{{\vec k}} 
\newcommand{\vK}{{\vec K}}  
\newcommand{\vb}{{\vec b}}  
\newcommand{\vp}{{\vec p}}  
\newcommand{\vq}{{\vec q}}  
\newcommand{\vQ}{{\vec Q}} 
\newcommand{\vx}{{\vec x}} 
\newcommand{\vh}{{\hat{v}}} 
\newcommand{\tr}{{{\rm Tr}}}  
\newcommand{\be}{\begin{equation}} 
\newcommand{\ee}{\end{equation}}  
\newcommand{\half}{{\textstyle\frac{1}{2}}}  
\newcommand{\gton}{\stackrel{>}{\sim}} 
\newcommand{\lton}{\mathrel{\lower.9ex \hbox{$\stackrel{\displaystyle <}{\sim}$}}}  
\newcommand{\ben}{\begin{enumerate}}  
\newcommand{\een}{\end{enumerate}} 
\newcommand{\bit}{\begin{itemize}}  
\newcommand{\eit}{\end{itemize}} 
\newcommand{\bc}{\begin{center}}  
\newcommand{\ec}{\end{center}} 
\newcommand{\bea}{\begin{eqnarray}}  
\newcommand{\eea}{\end{eqnarray}} 

\newcommand{\raae}{R_{AA}^e}

\newcommand{\comment}[1]{}
\newcommand{\ttbs}{\char'134}
\newcommand{\srt}{\mbox{$\sqrt{s}$}}
\def\GeV{{\rm GeV}}
\def\AGeV{{\rm A GeV}}
\def\GeVc{{\rm GeV/c}}
\def\AGeVc{{\rm A GeV/c}}
\def\MeV{{\rm MeV}}
\def\fm{{\rm fm}}
\newcommand{\bold}[1]{\mbox{\boldmath $#1$}}    
\newcommand{\pp}{\bold p}
\newcommand{\Bbar}{\mbox{$\bar{\rm B}$}}
\newcommand{\etal}{{\it et al.}}
\newcommand{\gsim}{\mbox{\raisebox{-0.6ex}{$\stackrel{>}{\sim}$}}\:}
\newcommand{\lsim}{\mbox{\raisebox{-0.6ex}{$\stackrel{<}{\sim}$}}\:}
 
\newcommand{\eq}[1]{Eq.~(\ref{#1})}
\newcommand{\fig}[1]{Fig.~\ref{#1}}
\newcommand{\eff}{ef\!f}
\newcommand{\alphas}{\alpha_s}

\title{Elastic, Inelastic, and Path Length Fluctuations in Jet Tomography}

\date{\today  \hspace{1ex}}
 
\author{Simon Wicks}
\affiliation{Department of Physics, Columbia University, 
             538 West 120-th Street, New York, NY 10027}

\author{William Horowitz}
\affiliation{Department of Physics, Columbia University, 
             538 West 120-th Street, New York, NY 10027}
 
\author{Magdalena Djordjevic}
\affiliation{Department of Physics, Columbia University, 
             538 West 120-th Street, New York, NY 10027}
\affiliation{Department of Physics, The Ohio State University, 
             191 West Woodruff Avenue, Columbus, OH 43210}

\author{Miklos Gyulassy}
\affiliation{Department of Physics, Columbia University, 
             538 West 120-th Street, New York, NY 10027}

\begin{abstract} 
Jet quenching 
theory using perturbative QCD is extended to include (1) elastic as well as (2) inelastic parton 
energy losses and (3) jet path length fluctuations. The extended theory is applied to non-photonic single electron production in
central Au+Au collisions at $\sqrt{s} = 200$ AGeV.
The three effects combine to significantly reduce the discrepancy between theory and
current data without violating the global entropy bounds from multiplicity
and elliptic flow data. We also check for consistency with the pion 
suppression data out to 20 GeV. Fluctuations of the jet path 
lengths in realisitic geometry and the difference between the widths of fluctuations of elastic 
and inelastic energy loss are essential to take into account.
\end{abstract}


\maketitle 

\begin{figure}[!hbt] 
\hspace*{-0.2cm }
\epsfig{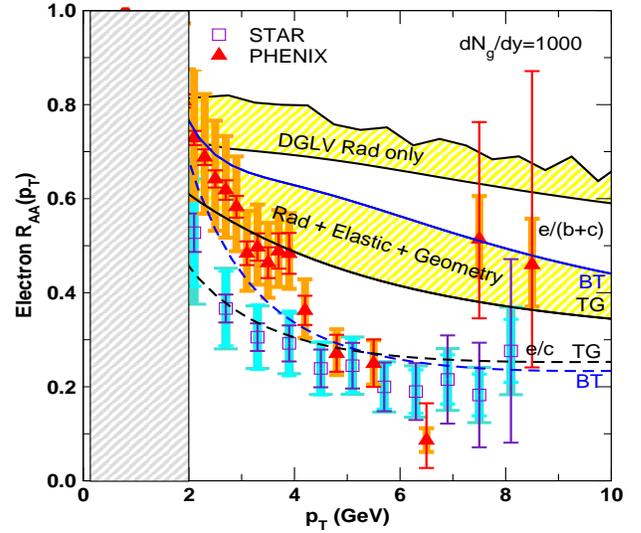} 
\begin{minipage}[t]{7.8cm}  
\vspace*{-0.6cm } 
\caption{\label{fig:eRAA} 
The suppression factor, $R_{AA}(p_T)$, of non-photonic electrons
  from decay of quenched heavy quark (c+b) jets is compared to
  PHENIX~\cite{Adare:2006nq} and STAR~\cite{elecQM05_STAR} data in central Au+Au reactions at
  200~AGeV.  Shaded bars indicate systematic errors, while
  thin error lines indicate statistical ones.
All calculations assume initial $dN_g/dy=1000$.
  The upper yellow band from~\protect{\cite{Djordjevic:2005db}}
  takes into account radiative energy loss only, using a fixed $L=6$ fm; the lower yellow band is our new prediction, including both
  elastic and inelastic energy losses as well as jet path length fluctuations. 
The bands provide a rough estimate of uncertainties from the leading log approximation for elastic energy loss. The dashed curves illustrate the lower extreme
of the uncertainty from production, by showing the electron suppression after both inelastic and elastic energy loss with bottom quark jets neglected.
}
\end{minipage}
\end{figure}

Light quark and gluon jet quenching observed via $\pi,\eta$
suppression~\cite{phenix_pi0} in Cu+Cu and Au+Au collisions at
$\sqrt{s} = 62-200$~AGeV at the Relativistic Heavy Ion Collider (RHIC)
has been remarkably consistent thus far with
predictions~\cite{Gyulassy:2003mc}-\cite{Vitev:2002pf}. However,
recent non-photonic single electron data~\cite{Adler:2005xv,Abelev:2006db,Adare:2006hc,elecQM05_STAR,Adare:2006nq} (which
present an indirect probe of heavy quark energy loss) have
significantly challenged the underlying assumptions of the jet
tomography theory (see~\cite{Djordjevic:2005db}).  A much larger
suppression of electrons than predicted was observed in the
$p_T\sim 4-8 $ GeV region
{(see \fig{fig:eRAA})}. These data falsify the assumption that heavy quark
quenching is dominated by radiative energy loss when the bulk QCD
matter parton density is constrained by the observed $dN/dy\approx
1000$ rapidity density of produced hadrons.

The observed ``perfect fluidity''~\cite{WhitePapers,BNLfluid} of the
sQGP at long wavelengths ($p_T < 2$ GeV) provides direct evidence
for highly nonperturbative bulk dynamics
~\cite{Gyulassy:2004zy,Hirano:2005wx}. Due to asymptotic
freedom, a breakdown of perfect fluidity and nonperturbative effects
are expected at $p_T$ several
times greater than the mean thermal energy, $3 T\sim 1-2$ GeV.
Prior to these electron data, pQCD based jet quenching theory
provided increasingly
reliable predictions  above
$p_T>5-7$ GeV~\cite{winter,WhitePapers} for the nuclear
modification of light parton jets 
~\cite{Gyulassy:2003mc}-\cite{Vitev:2002pf}. However, the non-photonic single electron data however raise the question of 
whether the novel nonperturbative physics of the strongly interacting Quark
Gluon Plasma (sQGP)~\cite{Gyulassy:2004zy} produced at RHIC 
could persist down to much smaller wavelengths
than previously expected.  This question is also of pragmatic
importance because high $p_T$ jets can be utilized as calibrated ``external''
tomographic probes of the bulk sQGP matter only if their dynamics can
be predicted reliably.

\begin{figure}[t] 
\hspace*{-0.2cm}
\epsfig{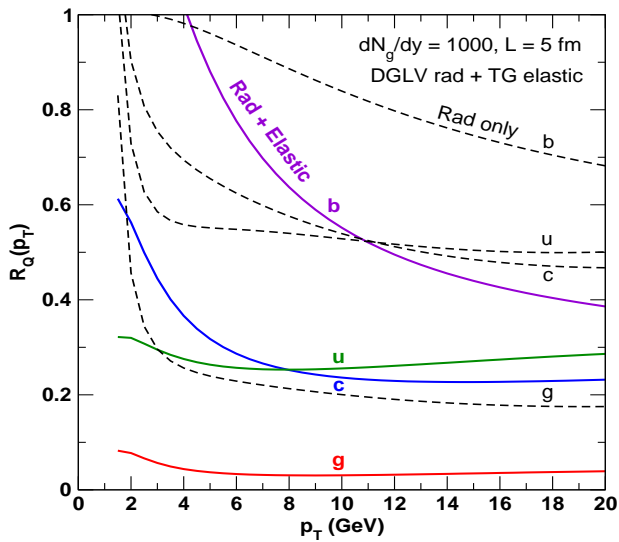} 
\begin{minipage}[t]{7.8cm}  
\vspace*{-0.6cm} 
\caption{\label{fig:ptDist2} 
Partonic nuclear modification, $R_{Q}^{II}(p_T)$ via Eq.(\ref{power2}),
for $g,u,c,b$ as a function
of $p_T$ for fixed L=5 fm path length and $dN_g/dy=1000$.
 Dashed curves
include only radiative energy loss,
while solid curves include elastic energy loss as well. 
}
\end{minipage} 
\end{figure}

The upper band of \fig{fig:eRAA} shows that the 
predictions from~\cite{Djordjevic:2005db} 
considerably underestimate the electron nuclear modification of data even out
to $p_T\sim 8$ GeV.  This discrepancy points to one or more of (1) missing
perturbative QCD physics, (2) incomplete understanding of the initial
heavy quark production and/or (3) novel non-perturbative mechanisms
affecting partonic physics out to $p_T > 10$ GeV. We note that
$p_T\sim 8$ GeV (single non-photonic) electrons originate in our
calculations from the fragmentation and decay of both charm and bottom
quarks with transverse momenta $p_T\sim 12\pm 4$ GeV (see \fig{fig:ptDist2}
in~\cite{Djordjevic:2005db}).

Possibility (3) is of course the most radical and would imply the
persistence of non-perturbative physics in the sQGP down to extremely
short wavelengths. Processes can be
postulated to improve the fit to the data~\cite{Rapp:2005at}, but at
the price of losing theoretical control of the tomographic information
from jet quenching data. DGVW~\cite{Djordjevic:2005db} showed that by
arbitrarily increasing the initial sQGP densities to unphysical $dN_g/dy
\gsim 4000$, the non-photonic electrons from heavy quarks can be
artificially suppressed to $R_{AA}\sim 0.5 \pm 0.1$.  Thus, to approach the electron data, conventional
radiative energy loss requires either a violation of bulk entropy bounds
or nonperturbatively large $\alphas$ extrapolations of the theory.
Even by ignoring the bottom contribution,
Ref.~\cite{Armesto:2005iq} found that a similarly excessive transport
coefficient~\cite{Baier:2002tc}, $\hat{q}_{\eff}\sim 14$ GeV$^2$/fm,
was necessary to approach the level of suppression of electrons in the data.

Bottom quark jets are very weakly quenched by radiative
energy loss. Using the FONLL production cross-sections, their contribution significantly reduces the single electron suppression~\cite{Djordjevic:2005db} compared to that of the charm jets alone. The ratio $R_{AA}$ is not sensitive to the scaling of all cross-sections by a constant. However, it is sensitive to any uncertainty in the relative contribution of charm and bottom jets to the electrons~\cite{Armesto:2005mz}. Recent data from STAR on electrons from p+p collisions~\cite{Abelev:2006db} may indicate an even larger uncertainty in the production than expected from FONLL. However, PHENIX p+p to electron data are compatible with the upper limit of FONLL predictions~\cite{Cacciari:2005rk,MNR}, similar to the comparison between FONLL and Tevatron data.

The discrepancy between the `DGLV Rad only' predictions and the data in \fig{fig:eRAA} and recent work~\cite{Mustafa:2004dr,Dutt-Mazumder:2004xk,Zapp:2005kt} motivated us to
revisit the assumption that pQCD elastic energy
loss~\cite{Bjorken:1982tu} is negligible compared to radiative. In
earlier studies, the elastic energy
loss~\cite{Bjorken:1982tu,Thoma:1990fm,Braaten:1991jj,Wang:1994fx,Mustafa:1997pm,Lin:1997cn}
was found to be $dE^{el}/dx\sim 0.3-0.5 $ GeV/fm, which was
erroneously considered to be small compared to the several GeV/fm expected
from radiative energy loss.  The apparent weakness of conventional
pQCD collisional energy loss mechanisms was also supported by parton
transport theory results~\cite{Molnar:2001ux}-\cite{Moore:2004tg},
which showed that the typical thermal pQCD elastic cross section,
$\sigma_{el}\sim 3 $mb, is too small to explain the differential
elliptic flow at high $p_T> 2 $ GeV and also underestimates the high
$p_T$ quenching of pions.

\begin{figure}[t] 
\hspace*{-0.2cm}
\epsfig{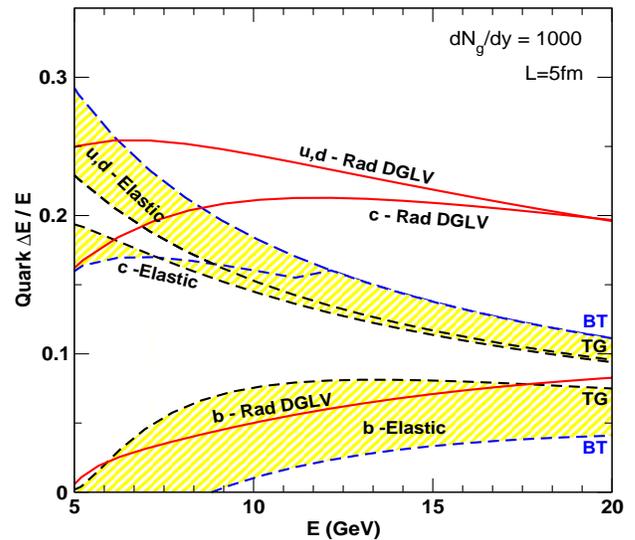} 
\begin{minipage}[t]{7.8cm}  
\vspace*{-0.6cm} 
\caption{\label{fig:ptDist} 
  Average $\Delta E/E$ for $u,c,b$ quarks as a function of $E$. A Bjorken
  expanding QGP with path length $L=5$~fm and initial density fixed
  by $dN_g/dy=1000$ is assumed. The curves are computed with the
  coupling $\alphas = 0.3$ held fixed. For Debye mass $\mu_D\propto(dN_g/dy)^{(1/3)}$,
    the gluon mass is $\mu_D/\surd 2$, the light quark mass is
    $\mu_D/2$, the charm mass is $1.2$ GeV, and
    the bottom mass is $4.75$ GeV.
  Radiative DGLV first order energy loss is compared to elastic parton
  energy loss (in TG or BT approximations). The yellow bands provide
  an indication of theoretical uncertainties in the leading log approximation to the elastic
  energy loss.}
\end{minipage} 
\end{figure} 

In contrast, Mustafa~\cite{Mustafa:2004dr} found that radiative and 
elastic average energy losses for heavy quarks were in fact comparable over a very 
wide kinematic range accessible at RHIC. In \fig{fig:ptDist}, we confirm 
Mustafa's finding and extend it to the light quark 
sector as well. The fractional energy loss,
$\Delta E/E$, from DGLV radiative for $u,c,b$ quarks (solid
curves; see also App. IB) is compared to  TG~\cite{Thoma:1990fm} and 
BT~\cite{Braaten:1991jj} estimates of elastic (dashed
curves; see also App. IA). For light quarks, the elastic energy loss 
decreases more rapidly with energy than radiative energy loss, but even at 20 GeV the elastic
is only 50\% smaller than the radiative. 

From \fig{fig:ptDist} we see that for $E>10$ GeV light and charm quark jets have 
elastic energy losses smaller but of the same order of magnitude as the inelastic losses. But due to the 
large mass effect~\cite{Dead-cone}-\cite{Zhang:2003wk},\cite{Armesto:2005iq}, 
both radiative and elastic energy losses remain significantly smaller for
bottom quarks than for light and charm quarks, but the elastic energy loss can now be greater than inelastic up to $\sim15$GeV. 
We present both TG and BT as a measure of the theoretical uncertainties of the Coulomb log (see App IC for
benchmark numerical examples). These are largest for the heaviest b quark. As they are not ultrarelativistic, the leading log approximation~\cite{Thoma:1990fm,Braaten:1991jj} breaks down in the kinematic range accessible at RHIC.
More rigorous computations of elastic energy loss~\cite{Djordjevic:2006tw} and numerical covariant transport techniques~\cite{Molnar:2001ux}
can be used to reduce the theoretical uncertainties in the elastic energy loss effects.

{\em Theoretical Framework.}\\
The quenched spectra of partons, hadrons, and leptons are calculated
as in~\cite{Djordjevic:2005db} from the generic pQCD convolution
\begin{eqnarray}
\frac{E d^3\sigma(e)}{dp^3} &=& \frac{E_i d^3\sigma(Q)}{dp^3_i}
 \otimes
{P(E_i \rightarrow E_f )}\nonumber \\
&\otimes& D(Q \to H_Q) \otimes f(H_Q \to e), \; 
\label{schem} \end{eqnarray}
where $Q$ denotes quarks and gluons.
For charm and bottom, the initial quark spectrum, 
$E d^3\sigma(Q)/dp^3$, is computed at next-to-leading order
using the code from~\cite{Cacciari:2005rk,MNR};
for gluons and light quarks, the initial distributions are computed 
at leading order as in \cite{Vitev:2002pf}. $P(E_i \rightarrow E_f )$ is the energy loss probability, $D(Q \to
H_Q)$ is the fragmentation function of quark $Q$ to hadron $H_Q$, and
$f(H_Q \to e)$ is the decay function of hadron $H_Q$ into the observed
single electron. We use the same mass and factorization scales as
in~\cite{Vogt} and employ the CTEQ5M parton densities~\cite{Lai:1999wy}
with no intrinsic $k_T$. As in~\cite{Vogt} we neglect shadowing of
the nuclear parton distribution in this application.

We assume that the final quenched energy, $E_f$, is large enough that 
the Eikonal approximation can be employed. We also assume that in Au+Au 
collisions, the jet fragmentation function into hadrons is the same as in 
$e^+e^-$ collisions. This is expected to be valid in the deconfined 
medium case, where hadronization of $Q\rightarrow H_Q$ cannot occur 
until the quark emerges from the sQGP. 

The main difference between our previous 
calculation~\cite{Djordjevic:2005db} 
and the present one is the inclusion of two new physics components in the  
energy loss probability $P(E_i \rightarrow E_f )$.
First, $P(E_i \rightarrow E_f)$ is generalized to include for the first time 
both elastic and inelastic energy loss and their fluctuations. We note that Vitev \cite{Vitev:2003xu} was the first to generalize the GLV formalism
to include {\it initial state} elastic energy loss effects
in d+Au. In this work, Eq. (\ref{fullconv}) extends the formalism to include
{\it final state} elastic energy loss effects  in $A+A$.

The second major
change relative to our previous applications is that we now take into account 
geometric path length fluctuations as follows:
\begin{eqnarray}
P(&E_i& \rightarrow E_i-\Delta_{rad}-\Delta_{el})=  \nonumber \\
&\;&\int\frac{d\phi}{2\pi} \int \frac{d^2 \vx_\perp}{N_{bin}(b)}  T_{AA}(\vx_\perp,\vb) \otimes  \nonumber \\
&P_{rad}&(\Delta_{rad}; L(\vx_\perp,\phi)) \otimes P_{el}(\Delta_{el}; L(\vx_\perp,\phi)).
\label{fullconv}\end{eqnarray}
Here 
\begin{eqnarray}
L(\vx_\perp,\phi)= \int d\tau 
\rho_p(\vx_\perp+\tau\hat{n}(\phi))/\langle\rho_p\rangle
\; \label{Leff}\end{eqnarray}
is the locally determined effective path length of the jet given its initial 
production point $\vx_\perp$ and its initial azimuthal direction $\phi$
relative to the impact parameter plane $(x,y)$. The geometric path 
averaging used here is similar to that used in~\cite{Gyulassy:2000gk}
and by Eskola et al.~\cite{Eskola:2004cr}. However, 
the inclusion of elastic energy losses together with path fluctuations
in more realistic geometries was not considered before.

\begin{figure}[t] 
\hspace*{-0.2cm }
\epsfig{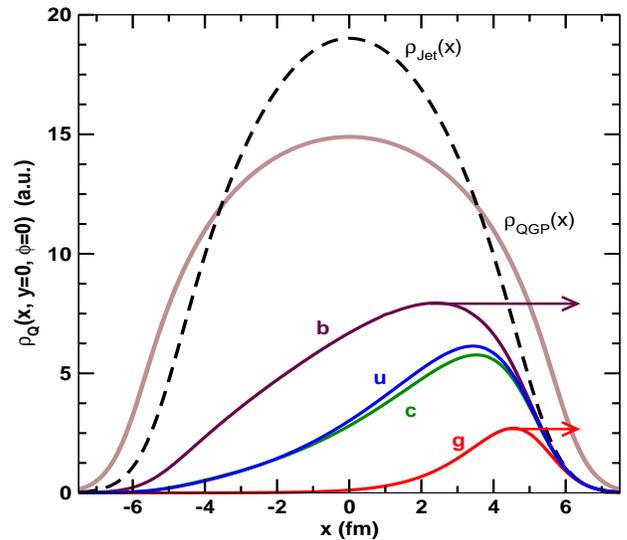} 
\begin{minipage}[t]{7.8cm}
\vspace*{-0.6cm}
\caption{\label{fig:Survival} Transverse coordinate
$(x,0)$ distribution of surviving
$p_T=15$ GeV, $Q=g,u,c,b$ jets moving in direction $\phi=0$
as indicated by the arrows. Units are arbitrary for illustration.
The transverse (binary collision) distribution
of initial jet production points, $\rho_{\rm Jet}(x,0)$, is shown
at midrapidity for Au+Au collisions at $b=2.1$ fm.
The ratio $\rho_Q/\rho_{\rm Jet}$ (see Eq.(\ref{rhoQ}))
gives the local quenching factor
including elastic and inelastic energy loss though
the bulk QGP matter distributed as $\rho_{\rm QGP}(x,0)$.}
\end{minipage}
\end{figure}

We consider a diffuse Woods-Saxon nuclear density profile~\cite{Hahn}, which creates a participant transverse density,
$\rho_p(\vx_\perp)$, computed using the Glauber profiles,
$T_A(\vx)$, with inelastic cross section $\sigma_{NN}=42$ mb. The
bulk sQGP transverse density is assumed to be proportional to this
participant density, and its form is shown (for the $y=0$
slice) in \fig{fig:Survival} by the curve labeled $\rho_{{\rm QGP}}$. The distribution
of initial hard jet production points, $\vx_\perp$, is assumed on the
other hand to be proportional to the binary collision density,
$T_{AA}=T_A(\vx+\vb/2)T_A(\vx-\vb/2)$. This is illustrated in \fig{fig:Survival} by the narrower curve labeled $\rho_{\rm Jet}$.

The combination of fluctuating DGLV radiative~\cite{Djordjevic:2003zk}
with the new elastic energy losses and fluctuating path lengths (via 
the extra $d^2\vx_\perp d\phi$ integrations) adds a high
computational cost to the extended theory specified by
Eqs.~(\ref{schem},\ref{fullconv}). In this first study with the extended 
theory, we explore the relative order of magnitude of the competing effects
by combining two simpler approaches.

In approach I, we parameterize the heavy quark pQCD spectra by a simpler 
power law, $E d^3\sigma_Q/d^3k \propto 1/p_T^{n+2}$, with a slowly 
varying logarithmic index $n\equiv n(p_T)$. For the pure power
law case, the {\em partonic}  modification factor, 
$R_Q=d\sigma_Q^{final}/d\sigma_Q^{initial}$, (prior to fragmentation) 
is greatly simplified. This enables us to perform the path length 
fluctuations numerically via
\begin{eqnarray}
R^I_Q = \int\frac{d\phi}{2\pi}\int \frac{d^2 \vx_\perp}{N_{bin}(b)} 
\; T_{AA}(\vx_\perp,\vb)\nonumber\\
\int d\epsilon (1-\epsilon)^n P_Q^I(\epsilon; L(\vx_\perp,\phi)),
\label{power}
\end{eqnarray}
where
\be
P^I_Q(\epsilon; L) = \int dx P_{Q,rad}(x; L) P_{Q,el}(\epsilon-x; L).
\label{peps}
\ee
Both the mean and width of those fractional 
energy losses depend on the local path length. 
(See App ID for numerical illustrations of Eq.(\ref{peps}) for a fixed 
$L=5$ fm light quark case.)

\begin{figure}[t] 
\hspace*{-0.2cm }
\epsfig{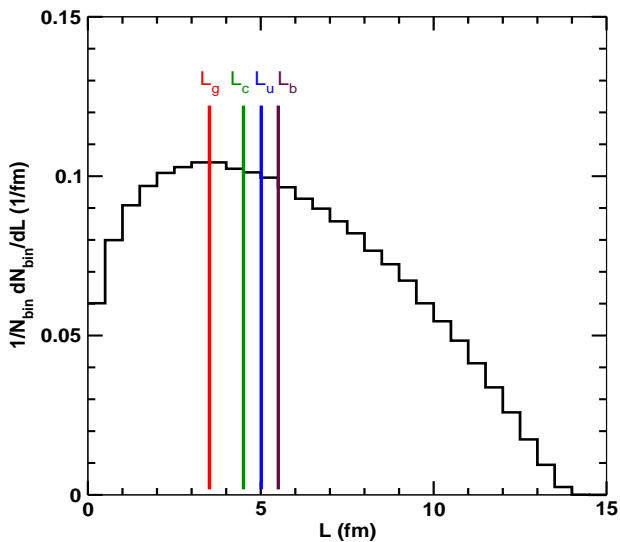} 
\begin{minipage}[t]{7.8cm}
\vspace*{-0.6cm}
\caption{\label{fig:pofl} Distribution of path lengths (given by \eq{Leff}) traversed by hard scatterers in 0-5\% most central collisions; the lengths, $L(\vx_\perp,\phi)$, are weighted by the probability of production and averaged over azimuth. An equivalent formulation of \eq{power} is $R_Q^I=\int dL 1/N_{bin} dN_{bin}/dL \int d\epsilon (1-\epsilon)^n P_Q^I(\epsilon;L)$. Since the distribution $1/N_{bin} dN_{bin}/dL$ is a purely geometric quantity, it is the same for all jet varieties. Also displayed are the single, representative pathlengths, $L_Q$, used as input in approach II. Note the hierarchy of scales with glue requiring the shortest, then charm, light quarks, and bottom the longest effective pathlength.}
\end{minipage}
\end{figure}

We emphasize, however, 
that no externally specified {\it a priori} path length, $L$, appears in 
Eq.~(\ref{power}); the path lengths are allowed to 
explore the whole geometry. \fig{fig:pofl} shows the broad distribution of lengths probed by hard partons in approach I.

In the second approach, we determine effective path lengths, $L_Q$, 
for each parton flavor, $Q$, by varying fixed $L$ predictions 
${R}^{II}_Q(p_T, L)$ and comparing to approach I; see \fig{fig:pofl}. In approach II,  
${R}^{II}_Q(p_T, L)$ is calculated directly from Eq.~(1) with 
$P(E_i \rightarrow E_f)$ in Eq.~(2) replaced by the fixed $L$ approximation
\begin{eqnarray}
 P(E_i \rightarrow E_i - \Delta_{el} - &\Delta_{rad}&, L)\approx \nonumber\\
 P_{rad}(&\Delta_{rad}&; L)\otimes P_{el}(\Delta_{el}; L).
\label{fixedL}\end{eqnarray}
Here, jet quenching is performed via two independent branching 
processes in contrast to the additive convolution in Eq.~(\ref{power}).
For small energy losses the two approaches are similar. They differ
however in the case of long path lengths when the probability of complete
stopping approaches unity. In the convolution method,
the probability of $\epsilon>1$ is interpreted as complete stopping,
whereas in the branching algorithm the long path length case is just highly
suppressed. In both cases we take into account the
small finite probability that the fractional energy loss $\epsilon \leq 0$ due to fluctuations.

To illustrate the difference in approach II, consider the case of power law
initial $Q$ distributions as in Eq.~(\ref{power}). In this case
\begin{eqnarray}
R^{II}_Q(p_T,L_Q)
&\equiv& \langle(1-\epsilon_{Q}^r(L_{Q}))^n (1-\epsilon_{Q}^e(L_{Q}))^n \rangle_{\Delta E}.
\nonumber \\
\;\label{power2}\end{eqnarray}
The branching implementation is seen via the product of two distinct 
factors in contrast to the one quenching factor in Eq.~(\ref{power}).
For small $\langle \epsilon_Q^{r,e} \rangle $ both approaches 
obviously give rise to the same $R_Q=1- n \langle \epsilon_Q\rangle$.

Due to the high computational cost in approach I, only the TG elastic is used for the heavy quarks and only BT for light quarks. The Coulomb log uncertainties are estimated only in approach II.

In both approaches, fluctuations of the radiative energy loss due to gluon number fluctuations are computed as discussed in detail in
Ref.~\cite{Djordjevic:2005db,Djordjevic:2004nq}. This involves using
the DGLV generalization~\cite{Djordjevic:2003zk} of the GLV opacity
expansion~\cite{Gyulassy:2000er} to heavy quarks. Bjorken longitudinal expansion is taken into account by evaluating the bulk density at an average time $\tau = L/2$~\cite{Djordjevic:2005db,Djordjevic:2004nq}. For elastic energy loss, the full 
fluctuation spectrum is approximated here by a Gaussian centered at the average energy loss with
variance $\sigma_{el}^2 = 2 T \langle \Delta E^{el}(p_T,L)\rangle $
\cite{Moore:2004tg}. In approach I the correct, numerically intensive integration through the Bjorken expanding medium provides $\Delta E^{el}(p_T,L)$. In approach II the $\tau = L/2$ approximation is again used; numerical comparisons show that for $L\sim2-7$ fm this reproduces the full calculation well. Finally, we note that we use the additional numerical simplification of keeping the strong coupling constant $\alphas$ fixed at $0.3$.

{\em Numerical Results: Parton Level} \\
In \fig{fig:ptDist2}, we show the quenching pattern of $Q$ from the second approach
for a ``typical'' path length scale $L=5$ fm, similar to that used in previous calculations~\cite{Djordjevic:2005db}. 
The curves show $R_Q(p_T)$, prior to hadronization, for
$Q=g,u,c,b$. The dashed curves show the quenching arising from only the 
DGLV radiative energy loss. The solid curves show the full results
after including TG elastic as well as DGLV radiative energy loss. Adding 
elastic energy loss is seen to increase the quenching of all flavors for fixed 
path length. Note especially the strong nonlinear 
increase of the gluon suppression and the factor $\sim 2$ 
increase of the bottom suppression. The curious switch of 
the $u$ and the $c$ quenching reflects the
extra valence (smaller index $n_u$) contribution to light quarks.

\fig{fig:ptDist2} emphasizes the unavoidable result of using a fixed, ``typical'' 
path length scale, $L$, in jet tomography: the pion 
and single electron quenching can never be similar. If pions were produced only by light quarks and electrons only by charm, 
then we would expect comparable quenching
for both. However, contributions from highly quenched gluons 
decrease the pion $R_{AA}$ while weakly quenched bottom quarks 
increase the electron $R_{AA}$. Therefore, in the fixed length 
scenario, we expect a noticeable difference between pion 
and single electron suppression patterns.

The solid curves of \fig{fig:Survival} labeled by the parton flavor $Q=g,u,c,b$
show the relative transverse coordinate density of surviving jets defined by
\begin{eqnarray}
\rho_Q(\vx,\phi)\equiv \rho_{\mathrm{Jet}}(\vx)\int d\epsilon (1-\epsilon)^n P_Q^I(\epsilon; L(\vx, \phi)).
\label{rhoQ}
\end{eqnarray}
$\rho_Q$ is given by the initial transverse $\vx$ production distribution 
times the quenching factor from that position in direction $\phi$
with final momentum $p_T$. The case shown is for a $p_T=15$~GeV 
jet produced initially at $(x,0)$ and moving in the direction $\phi=0$
along the positive x axis. The quenching is determined by the participant 
bulk matter along its path $\rho_{\rm QGP}(x+ vt,0,t)$, and varies with $x$ because the local path length $L=L(x,0,0)$ changes according to Eq.~(\ref{Leff}).

What is most striking in \fig{fig:Survival} is the hierarchy of $Q$-dependent length 
scales. No single, representative path length can account for the distribution of all 
flavors. In general heavier flavors are less biased toward the surface 
(in direction $\phi$) than lighter flavors since the energy loss decreases 
with the parton mass. Gluons are more surface biased than light 
quarks due to their color Casimir enhanced energy loss. In addition, note
the surprising reversal of the $u$ and $ c$ suppressions, also seen in \fig{fig:ptDist2}. 
\fig{fig:ptDist} shows that the energy loss of $c$ is somewhat less than for $u$; 
however, the higher $p_T$ power index, $n$,
of $c$ relative to $u$ -- as predicted by pQCD and due to 
the valence component of $u$ -- compensates by amplifying its quenching. 

However, none of the distributions can be
categorized as surface emission. The characteristic widths of these
distributions range from $\Delta x \approx 3-6$ fm. We show below that such a large
dynamic range of path length fluctuations is essential for
consistent reproduction of both electron and pion data.

\begin{figure}[b] 
\hspace*{-0.2cm}
\epsfig{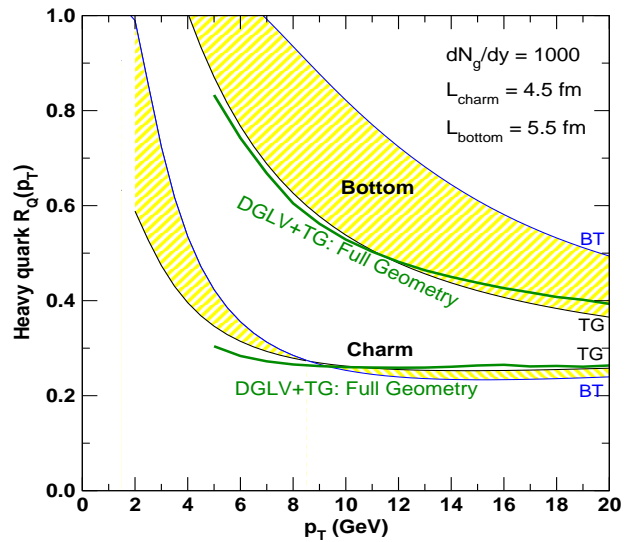} 
\caption{\label{fig:cbRAA} Heavy quark jet quenching before fragmentation into mesons for $dN_g/dy=1000$. Solid green curves show the results of approach I based on Eq.~(\ref{power}) including full geometric path length fluctuations and DGLV radiative and TG elastic energy loss for c and b quarks. Upper and lower 
yellow bands show predictions using approach II via Eq.~(\ref{power2}) with effective path lengths taken as $L_{c}= 4.5$ and $L_{b}= 5.5$ fm. As previously noted in \fig{fig:ptDist}, the difference between TG and BT curves indicates an estimate of the magnitude of theoretical uncertainties in the elastic energy loss.}
\end{figure}

\begin{figure}[!th] 
\hspace*{-0.8cm }
\epsfig{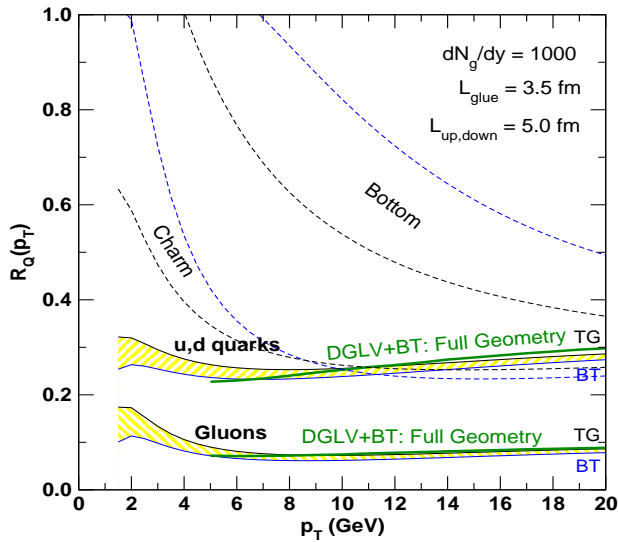} 
\begin{minipage}[thb]{7.8cm}  
\vspace*{-0.2cm } 
\caption{\label{fig:qRAA} As in \fig{fig:cbRAA} 
but for light $u,d$ quarks and gluons. The yellow bands
are computed in this case with 
effective $g,u$ path lengths $L_{g}= 3.5$ and 
$L_{u}= 5.0$ fm based on Eq.~(\ref{power2}).
Note that  charm and light quark quenching
are similar in this $p_T$ range. }
\end{minipage}
\end{figure}

We turn next to Figs.~\ref{fig:cbRAA} and~\ref{fig:qRAA}, which show the interplay between the dynamical
geometry seen in \fig{fig:Survival} and the elastic-enhanced quenching of
partons.  In Figs.~\ref{fig:cbRAA} and~\ref{fig:qRAA} the solid green curves labeled ``DGLV+TG/BT: Full Geometry''
are the results using approach I based on Eq.~(\ref{power}). The
curves labeled TG and BT are from approach II based on
Eq.~(\ref{power2}). The effective fixed $L_Q$ in II were taken to
match approximately the green curves in which full path length
fluctuations are taken into account. This procedure is not exact because
of the different numerical approximations involved, but the trends are
well reproduced. The $L_Q$ are determined only to $\sim 0.5$ fm accuracy, as this suffices for our purposes here.
We show the comparison between approaches I and II for heavy quarks in \fig{fig:cbRAA} using $L_{c}= 4.5$ and $L_{b}= 5.5$ fm and for gluons and light quarks in \fig{fig:qRAA} using  $L_{g}= 3.5$ and $L_{u}= 5.0$ fm; see \fig{fig:pofl} for a visual comparison of the input length distributions used. This hierarchy of $Q$-dependent length scales is in accord with that expected from \fig{fig:Survival}.

Note that in comparison to the fixed $L=5$ fm case in \fig{fig:ptDist2}, geometric fluctuations reduce gluon jet quenching in \fig{fig:qRAA} by a factor
$\sim 2$. Nevertheless, even with path length fluctuations the gluons are still
quenched by a factor of 10 when elastic energy loss is included in addition to 
radiative. 
  
The amplified role of elastic energy loss is due to its smaller width
for fluctuations relative to radiative fluctuations.  Even in
moderately opaque media with $L/\lambda\sim 10$, inelastic energy loss
fluctuations are large because only a few, 2-3, extra gluons are
radiated \cite{GLV_suppress}. Thus, gluon number fluctuations, $O(1/\surd
N_g)$ lead to a substantial reduction in the effect of radiative
energy loss. On the other hand, elastic energy loss
fluctuations are controlled by collision number fluctuations,
$O(\sqrt{\lambda/L})$, which are relatively small in comparison for a significant proportion of the length scales probed. Therefore,
fluctuations of the elastic energy loss do not dilute the suppression of the nuclear
modification factor as much as $N_g$ fluctuations. The
increase in the sensitivity of the final quenching level to the
opacity is a novel and useful byproduct of including the elastic
channel; see \fig{fig:width} in Appendix D. The inclusion of elastic energy loss significantly reduces the 
fragility of pure radiative quenching~\cite{Eskola:2004cr} and therefore increases 
the sensitivity of jet quenching to the opacity of the bulk medium~\cite{HWG}.

{\em Numerical Results: Pions and Electrons} \\  
We now return to \fig{fig:eRAA} to discuss the consequence of including
elastic energy loss of $c$ and $b$ quarks on the electron spectrum. The inclusion of the collisional energy loss
significantly improves the comparison between theory and the single electron data. That is, the lower yellow band can reach below $R_{AA}\sim
0.5$ in spite of keeping $dN_g/dy=1000$, consistent with measured multiplicity, and
using a conservative $\alphas=0.3$. A large source of the uncertainty represented
by the lower yellow band is the modest
but poorly determined elastic energy loss, $\Delta E/E\approx 0.0
-0.1$, of bottom quarks (see \fig{fig:ptDist}). There is additional uncertainty from the relative contributions to electrons from charm and bottom
jets. The dashed lines show an extreme version of this in which charm jets are the only source of electrons. If the charm to bottom ratio given by FONLL calculations is accurate, the current data suggests that even the combined radiative+elastic pQCD mechanism is not
sufficient to explain the single electron suppression.

\begin{figure}[!t] 
\hspace*{-0.8cm }
\epsfig{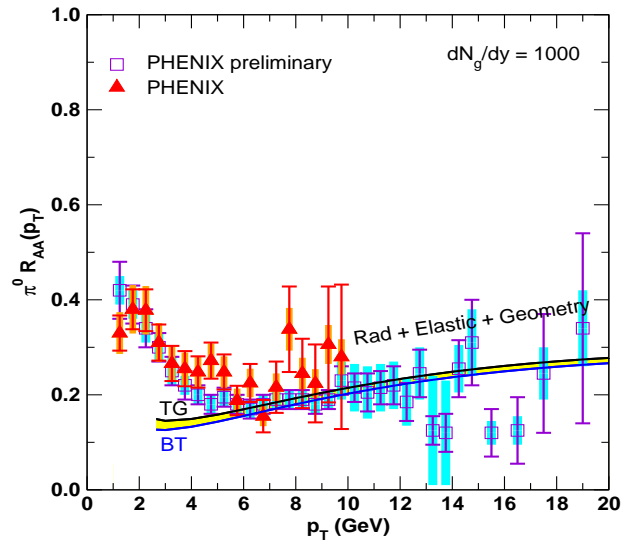} 
\begin{minipage}[thb]{7.8cm}  
\caption{\label{fig:piRAA} 
The consistency of the extended jet quenching theory
is tested by comparing its prediction to the nuclear modification
of the $\pi^0$ spectra observed by
PHENIX~\cite{phenix_pi0}.
}
\end{minipage}
\end{figure}

As emphasized in~\cite{Djordjevic:2005db}, any proposed energy loss mechanisms must also be checked for consistency with the extensive pion
quenching data~\cite{phenix_pi0}, for which preliminary data
now extend out to $p_T\sim 20$ GeV. 
This challenge is seen clearly in \fig{fig:ptDist2}, where for fixed $L=5$ fm, the addition of elastic energy loss 
would overpredict the quenching of pions. However, the simultaneous 
inclusion of path fluctuations leads to a decrease of the mean $g$ and
$u$,$d$ path lengths that partially 
offsets the increased energy loss. 
Therefore, the combined three effects considered here makes it 
possible to satisfy $R_{AA}^e<0.5 \pm 0.1$ without violating the bulk $dN_g/dy=1000$ 
entropy constraint and without violating the pion quenching constraint 
$R_{AA}^{\pi^0}\approx 0.2\pm 0.1$ now observed out to 20 GeV; see \fig{fig:piRAA}.
We note that the slow rise of $R_{AA}^{\pi^0}$ with $p_T$ in the present
calculation is due in part to the neglect of initial $k_T$ smearing
that raises the low $p_T$ region and the EMC effect that lowers
the high $p_T$ region (see~\cite{Vitev:2002pf}).

{\em Conclusions} \\
The elastic component of the energy loss cannot be neglected when considering pQCD jet quenching.
While the results presented in this paper are encouraging, further
improvements of the jet quenching theory will be required before
stronger conclusions can be drawn.

From an experimental
perspective, there is at present significant disagreement between measured
p+p to electron baselines~\cite{Abelev:2006db,Adare:2006hc}. In addition, direct measurement of $D$ spectra will be essential to
deconvolute the different bottom and charm jet quark dynamics. 

On the theoretical side, more work is needed to sort out
coherence and correlation effects between elastic and inelastic
processes that occur in a finite time and with multiple collisions.
Classical electrodynamics calculations presented
in~\cite{Peigne:2005rk} suggested that radiative and elastic processes
could destructively interfere over lengths far longer than previously
thought. As described in~\cite{Adil:2006ei},
a proper accounting of the current shows finite size effects persist
out only to the expected lengths of order the screening scale,
$1/\mu_D \lsim 1$ fm.  Additionally, work on the quantum mechanical treatment
of elastic energy loss in a finite
medium~\cite{Djordjevic:2006tw,Wang:2006qr} also concluded that
finite size
effects on $R_{AA}$ remain small except in peripheral collisions. 

There are several other open problems that require
further study. The inclusion of all the initial state effects from \cite{Vitev:2002pf} will be needed
to fully check the consistency of the pion predictions with the data. Only an approximate fluctuation spectrum
for elastic energy loss has been included here; still needed is an examination of the effect of the full fluctuation spectrum.

The radiative and elastic energy losses depend sensitively on the coupling,
$\Delta E^{rad}\propto\alphas^3$ and $\Delta
E^{el}\propto\alphas^2$. Future calculations
will have to relax the current fixed $\alphas$ approximation. In ~\cite{Peshier:2006hi}, the running of the coupling
is seen to increase the magnitude of the elastic energy loss and alter the energy dependence. More complete calculations of both radiative and elastic energy losses will involve integrals that probe momentum scales that are
certainly nonperturbative. Therefore it will be important to study the
irreducible uncertainty associated with the 
different maximum $\alphas$ cutoff prescriptions commonly used.

{\em Acknowledgments:} Valuable input from Azfar Adil on parton spectra
and discussions with Xin Dong, Barbara Jacak, John Harris, Peter L\'evai, Denes Molnar, Thomas Ullrich, Ivan Vitev, Xin-Nian Wang, 
and Nu Xu are gratefully 
acknowledged. This work is supported by the Director, Office of Science, 
Office of High Energy and Nuclear Physics, Division of Nuclear
Physics, of the U.S. Department of Energy under Grants No. DE-FG02-93ER40764.
M.~D.~acknowledges support by the U.S. Department of Energy under Grants 
No. DE-FG02-01ER41190 during the final parts of this work.  

\section{Appendix}

\subsection{Collisional Energy Loss}
The leading logrithmic expression for the elastic
energy loss of a jet with color Casimir $C_R$ 
in an ideal quark-gluon plasma with $n_f$ active quark flavors
and temperature, $T$,
is given by \cite{Bjorken:1982tu} 
\bea \frac{dE^{el}}{dx}=  C_R \pi \alphas^2 T^2 (1+\frac{n_f}{6}) f(v) \log(B_c)  \eea 
where the Coulomb log is controlled by
the ratio $B_c$ that involves
relevant minimum and maximum momentum transfers or impact parameters. 
For scattering in an assumed ultrarelativistic
$(m=0)$ gas of partons the jet velocity dependence is
\bea
f(v)=\frac{1}{v^2}\left(v +\half(v^2-1)\log(\frac{1+v}{1-v})\right)
\stackrel{v\rightarrow 1}{\longrightarrow} 1
\eea
Estimates for $B_c$ differ below asymptotic $(E\gg T)$
energies and are given in \cite{Bjorken:1982tu}, \cite{Thoma:1990fm}, and
\cite{Braaten:1991jj} that we denote by Bj, TG, and BT respectively: \bea
B_{\rm Bj}&= & \hspace{0.15in}\left( 4E_pT \right)/\left(\mu^2\right)  \nonumber \\[1ex]
B_{\rm TG} &=& \hspace{0.15in}\left( \frac{4pT}{(E_p-p+4T)} \right) / \left(\mu\right) \nonumber\\[1ex]
B_{\rm BT} &=& \left\{ \begin{array}{ll}
\left(2^{\frac{n_f}{6+ n_f}}\; 
    0.85 \;E_p T \right)/ \left(\frac{\mu^2}{3}\right) & 
E_p \gg \frac{M^2}{T} \\[2ex] 
\left(4^{\frac{n_f}{6+ n_f}}\; 
    0.36 \; \frac{(E_p T)^2}{M^2} \right)/ \left(\frac{\mu^2}{3}\right) & 
E_p\ll \frac{M^2}{T} 
\end{array}\right.
\nonumber \\[1ex]
\eea
with the crossover between $E_p \ll \frac{M^2}{T}$ and $E_p \gg \frac{M^2}{T}$ being taken at $E_p = \frac{2 M^2}{T}$ for numerical computation. 

\subsection{DGLV Radiative Energy Loss}

For completeness we also record the DGLV formula for radiative energy
loss used in our calculations. We neglect finite kinematic limits on the 
momentum transfer $q$ integral, and perform the finite $0<k_\perp
\le k_{max}=2 p x(1-x)$ and azimuthal $0\le \phi\le 2\pi$
integrations analytically. The mean fractional radiative energy loss
can be then evaluated numerically from the expression
 \bea
 \frac{\Delta E_{ind}^{(1)}}{E} &=& \frac{C_F \alphas}{\pi} 
\frac{L}{\lambda_g} \int^{1-\frac{M}{E_p+p}}_{\frac{m_g}{E_p+p}} dx \int^{\infty}_0 
\frac{4 \mu^2 q^3 dq}{\left( \frac{4Ex}{L} \right)^2 + (q^2 + \beta^2)^2} \nonumber \\
&\;& \;\;\hspace{0.5in} \times (A \log B + C)
 \eea
 where
\bea
\beta^2 &=& m^2_g (1-x) + M^2 x^2 \\
\lambda_g^{-1} &=& {\rho_g \sigma_{gg} + \rho_q \sigma_{qg}} \\
\sigma_{gg} &=& \frac{9\pi\alphas^2}{2 \mu^2} \mbox{ and } \sigma_{qg} = \frac{4}{9} \sigma_{gg}
\; \; .\eea

We employ the asymptotic 1-loop
transverse gluon mass $m_g = {\mu}/{\sqrt{2}}$.
The A,B,C functions denote
 \begin{eqnarray}
 A &=& \frac{2 \beta^2}{f_{\beta}^3} \left( \beta^2 + q^2 \right) \\ 
 B &=& \frac{(\beta^2 + \kmax) (\beta^2 \Qm + \Qp\Qp  + 
\Qp f_{\beta} )}{ 
\beta^2 (\beta^2 (\Qm - \kmax) - \Qm \kmax + {\Qp}\Qp + f_{\beta}
f_{\mu}) } 
\\
 C &=& \frac{1}{2 q^2 f_{\beta}^2 f_{\mu}}  
\left[ \beta^2 \mu^2 (2 q^2 - \mu^2) + \beta^2 (\beta^2 - \mu^2) \kmax \right.
\nonumber \\ 
&\;& + \Qp (\beta^4 - 2 q^2 \Qp) 
  +
   f_{\mu} 
(\beta^2 (-\beta^2 - 3 q^2 +\mu^2) \nonumber\\
&\;& \left. + 2 q^2 \Qp) + 3 \beta^2 q^2 \Qmk \right]
 \end{eqnarray}
where the abbreviated symbols denote
\bea
K&=&k_{max}^2 =2p x(1-x)\\
Q^\pm_\mu&=&q^2\pm \mu^2 \\
Q^\pm_k&=& q^2\pm k_{max}^2\\
f_\beta &=& f(\beta, Q^-_\mu,Q^+_\mu)\\
f_\mu &=&f(\mu,Q^+_k,Q^-_k)
\eea
with $f(x,y,z)=\sqrt{x^4+2x^2 y+ z^2}$. 

\subsection{Benchmark Numerical Examples}

In this section we record numerical benchmark cases
of both the elastic and radiative mean energy loss
to illustrate the above formulas.
Consider a uniform Bjorken cylinder with  density 
\begin{equation}
\rho(\tau) = \frac{1}{\pi R^2 \tau} \frac{dN}{dy} 
\end{equation}
We assume $R=6$ fm. 
The temperature evolves as
\[
T(\tau) = \left( \frac{\pi^2}{1.202} \frac{\rho(\tau)}{9 n_f + 16} \right)^{\frac{1}{3}}
\]
where $n_f$ is the number of active quark flavors.
The effective static approximation simulates
the effect of Bjorken expansion by evaluating $T$ at $\tau=L/2$,
 where $L$ is the jet path length 
to the the cylinder surface. 
The gluon density is computed from
$\rho_g = \frac{1.202}{\pi^2}\times 16 T^3$, 
and the density of quarks plus antiquarks is
$\rho_q = \frac{1.202}{\pi^2}\times 9 n_f T^3$.
The Debye mass squared is $\mu^2 = 4 \pi \alphas^2 T^2 (1+\frac{n_f}{6})$. 
In Table I the results for $n_f=0$ are given for 
a charm jet of energy $10\le E\le 15$.

\begin{table}
\begin{tabular}{c|c|c|c|c}
 & Radiative $\frac{\Delta E}{E}$& \multicolumn{3}{c}{Collisional $\frac{\Delta E}{E}$} \\
$E_{jet}$ (GeV) & DGLV & Bj & TG & BT \\
\cline{1-5}
10 & 0.2111 & 0.2022 & 0.1594 & 0.1596 \\
11 & 0.2126 & 0.1894 & 0.1506 & 0.1552 \\
12 & 0.2129 & 0.1782 & 0.1430 & 0.1621 \\
13 & 0.2123 & 0.1683 & 0.1358 & 0.1530 \\
14 & 0.2110 & 0.1596 & 0.1294 & 0.1450 \\
15 & 0.2093 & 0.1518 & 0.1237 & 0.1379
\end{tabular}
\caption{$\frac{\Delta E}{E}$ benchmark test cases for a charm jet 
($m=1.2$ GeV, $C_R = \frac{4}{3}$) with fixed $\alphas=0.3$. The path length is $L=5$ fm; $\hbar c = 0.197$ GeV fm. The density is $\frac{dN}{dy}=1000$ ($n_f=0$), giving $T=0.2403$ GeV, $\mu=0.4666$ GeV, and $\lambda_g=1.2465$ fm. 
For radiative, the q integration limits are taken
to be 0.0001 and 50.}
\end{table}

\subsection{Energy Loss Fluctuation Spectrum}

\begin{figure}[tb] 
\epsfig{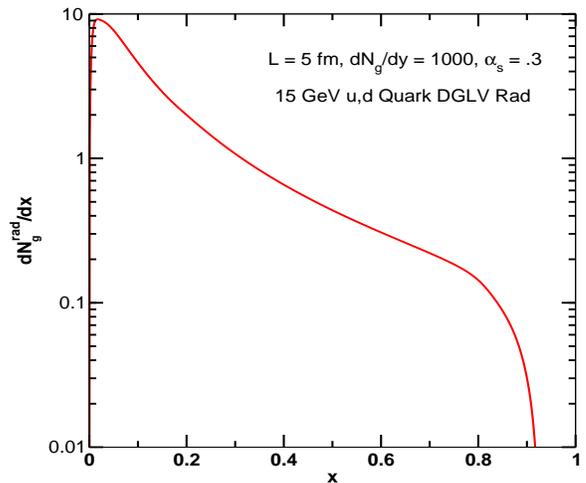} 
\begin{minipage}[thb]{7.8cm}  
\caption{\label{fig:dNdx} 
Example of the induced DGLV radiation gluon number spectrum for
a 15 GeV up quark, with path length $L=5$ fm, $dN_g/dy = 1000$, $\alphas = 0.3$, and $A_\perp$=118.7 fm$^2$.
}
\end{minipage}
\end{figure}

\begin{figure}[!hct] 
\vspace{0.3cm }
\epsfig{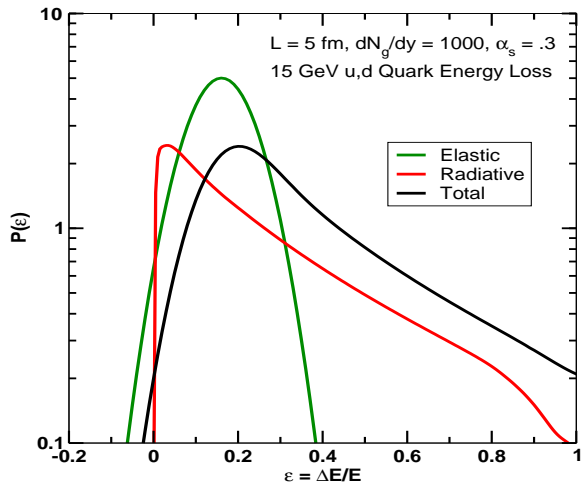} 
\begin{minipage}[thb]{7.8cm}  
\caption{\label{fig:PeDist} 
The distribution of fractional radiative energy
loss (or gain), $\epsilon=\Delta E/ E$, for the case
considered in \fig{fig:dNdx}. The narrow (green) curve corresponds to BT elastic energy loss fluctuations; based on the input from \fig{fig:dNdx} the lower, broader (red)
curve corresponds to inelastic energy loss due to
gluon number fluctuations.
The part of the radiation spectrum with $\epsilon>1$ is
replaced with $0.0213\delta (1-x)$, and there is a
contribution from zero gluon number fluctuations, $0.2377\delta(x)$, not shown. The continuous part
of the radiative
distribution has integrated norm 0.7374 with mean 0.2026.
The top (black) curve corresponds to 
the convolution of elastic and inelastic
energy fraction fluctuations.
}
\end{minipage}
\end{figure}

This section illustrates the fluctuation spectra of
induced gluon number and the distribution of fluctuating
energy loss for a specific case of a 15 GeV up quark jet
with path length $5$ fm. \fig{fig:dNdx} shows the first order
induced gluon number distribution $dN_g/dx$ for this case.
\fig{fig:PeDist} shows the fractional
radiative energy loss distribution taking into
account Poisson fluctuations of the gluon number 
computed as in \cite{GLV_suppress}. The finite probability, $P(n_g=0)=0.2377$, of radiating zero gluons contributes a $\delta(\epsilon)$ that is not 
shown. There is also a finite probablity, $0.0213$, of complete stopping with
$\epsilon>1$.

\begin{figure}[!hct] 
\vspace{0.3cm }
\epsfig{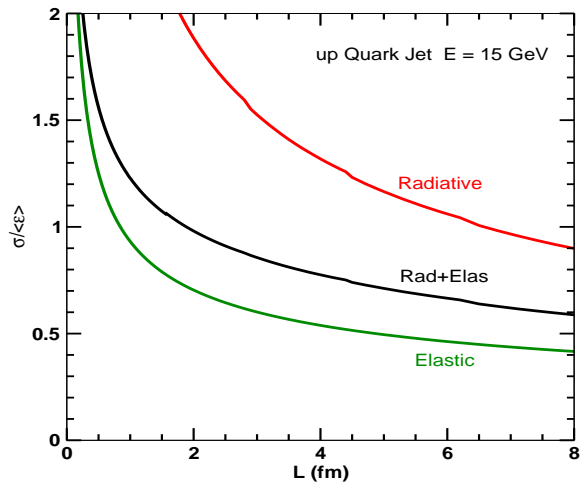} 
\begin{minipage}[thb]{7.8cm}  
\caption{\label{fig:width} 
The ratio of rms width, $\sigma(L)$, to the mean fraction
energy loss $ \langle \epsilon\rangle$ for radiative, elastic and convolved 
energy loss distributions is shown as a function of the path length, $L$,
for the Bjorken expanding plasma with $dN_g/dy=1000$.
The case of an up quark jet with $E=15$ GeV is shown.
Notice that the elastic distribution is significantly narrower
than the radiative one. This amplifies the effect of elastic energy loss on $R_{AA}$ relative to radiative.
}
\end{minipage}
\end{figure}

The width of the elastic energy fluctuations seen in \fig{fig:width}
is significantly smaller than the radiative width. The narrowing of the convoluted elastic 
plus radiative distributions significantly 
reduces the distortion effects due to fluctuations. Because of the steep $p_T$ fall off of the initial unquenched parton spectra, the smaller
width of the elastic energy fluctuations considerably amplifies 
the quenching effect due to collisional energy loss in comparison
to the larger but much broader radiative contribution. In terms
of an effective renormalization
$\langle \epsilon \rangle \rightarrow Z_{\eff} \langle \epsilon\rangle$
as discussed in \cite{GLV_suppress}, $Z_{\eff}$ is closer to unity
than the renormalization $Z_{rad}\sim 0.5$ characteristic 
of pure radiative energy loss distributions.


\end{document}